# The Development of Sealed UV Sensitive Gaseous Detectors and their Applications

L.Periale[1,2], V. Peskov[1], A. Braem[1], Di Mauro[1], P. Martinengo[1], P. Picchi[1,3], F. Pietropaolo[1,4], H. Sipila[4]

[1] CERN, Geneva, Switzerland
[2] INAF, Turin, Italy
[3] INFN, Frescati, Italy
[4] INFN, Padova, Italy
[5] Oxford Instrument, Expoo, Finland

**Abstract**

We have developed commercial prototypes of sealed gaseous detectors combined with CsI photocathodes and/or filled with photosensitive vapors. The rirst results of application of these devices for the detection of flames in daylight conditions and for the detection of scintillation lights from noble liquids will be presented. The main conclusion from our studies is that for some applications the sealed UV sensitive gaseous detectors have superior performance (higher practical quantum efficiency and better signal to noise ratio) than existing commercial UV sensitive detectors. Additionally, they are much cheaper.

1. **Introduction**

Photosensitive wire chambers (introduced by Seguinot et al., [1] and Bogomolov et al.,[2]) and gaseous detectors with CsI photocathodes ( first developed by Charpak et al., [3] and Dangendorf et al., [4]), operating in flushed by gas mode found widespread applications in experimental practice. For example, multiwire proportional chambers combined with CsI photocathode are under development and testing for several large - scale high –energy physics experiments, for example ALICE, HADES and others [5]. The main advantages of these detectors are their simplicity and large sensitive area (in some cases square meters). There were also some efforts to develop sealed gaseous detectors with CsI photocathode (see for example [6,7]) however for high- energy physic applications they had no special advantages over the flushed ones and for this reason did not receive any application in high-energy physics experiments yet.
However, there are other applications which require sealed UV sensitive detectors. Examples could be: for the detection of flames in daylight conditions [7], noble liquid TPC [8], spectroscopy, synchrotron radiation physics, biology [9], various security devices.

The main aim of this work is to investigate the possibility of manufacturing sealed UV sensitive gaseous detectors with CsI photocathode and/or filled with photosensitive vapors at a cost much lower than any other alternative detectors (for example vacuum photomultipliers) and evaluating of their efficiency, performance and long-term stability.
The abilities of these detectors will be demonstrated on two examples: detection of flames in daylight condition and detection of scintillation lights from noble liquids

## 2. Set -up for the Manufacturing and Testing of Sealed Gaseous Detectors

The- set up for the manufacturing of sealed gaseous detector is show schematically in Fig.1. Essentially, it consists of the detector to be sealed connected to the pump and gas systems. The detectors used in this work were stainless steal cylindrical single -wire counters (the diameters of the cathode cylinder of 20 and 30 mm, 110mm and 150mm in length, diameter of the anode wire was of 0.1mm) with quartz (diameter of the cathode of 20 mm) or $MgF_2$ windows (diameter of the cathode of 30 mm). The interface between the cathode and the anode wire was manufactured from ceramics. Two types of photosensitive detectors were manufactured: with CsI photocathode (see Fig. 2) and filled with photosensitive vapors. Gaseous detectors with CsI photocathodes were manufactured by the following procedure. First, before the window was mounted, an CsI layer was deposited on the inner part of the cylindrical cathode on the surface located opposite to the window. The coating with the CsI layer was done at the CERN CsI evaporation facility. After this the cylinder was extracted from the evaporation plant and the window was attached to the cathode cylinder and sealed thereafter. The total time during which the cathode was exposed to air was around one hour. The detector was then heated to 65°C and pumped to a vacuum of better than $10^{-6}$ Torr for several days. After such a step, it was cooled to room temperature and filled with one of the gas mixtures: $Ar+10\%CO_2$ or $Ar+10\%CH_4$ at a total pressure of 1 atm. In the case of the detector filled with the photosensitive vapor the window was sealed to the cathode cylinder from very beginning. The detector was then heated to 150°C and pumped to a vacuum of $10^{-6}$ for several days. After, the detector was cooled to room temperature and ethylferrocene (EF) vapors [10] were introduced. Finally, the detector was filled by $Ar+15\%CO_2$ at a total pressure of 1atm.
The quantum efficiency (QE) at various wavelengths was measured with the help of a monohromator (see Fig. 3). The absolute intensity of the light beam which exited the monochromator and entered the detector was calibrated in the wavelength interval of 140-200 nm with the help of an ionization chamber filled with TMAE vapors (the QE of TMAE is well known [11]) and in the wavelength interval of 200-300 nm with respect to the photodiodes calibrated by Hamamatsu.
The stability of the manufactured detector was monitored by using a pulse $D_2$ lamp (see Fig. 4) having a peak of light emission at 160 nm. The intensity of the pulsed $D_2$ lamp was sufficient enough to obtain a clear signal even at gain one and this allowed to measure accurately the gain vs. voltage and to monitor the detector efficiency.
In some control measurements a Hg lamp combined with narrow-band filters having a peak of transmission at 200 nm was also used.

## 3. The Results of Some Measurements

The results of some of our measurements are presented in Figs 5-7.
In Fig. 5 are shown the QE vs. wavelength measured for both type gaseous detectors: with CsI photocathodes and filled with EF vapors. Comparing the measured QE to the values presented in our previous work [13] one can conclude that the quantum efficiency of our CsI photocathodes was less by almost a factor of two. We attribute this to the fact that or photocathdes were exposed to air for long time - around 1 hour. In contrast, in work [13] the photocathdes were transferred to the detector within a few min. The value of the QE for the detectors filled with the EF was closed to the one calculated for the given length of the detector (150 mm) and the temperature of 22-25°C, which was in the laboratory during the measurements (the contribution of the EF absorbed layer was also taken in to the account [13]).
In Fig. 6 are presented the results of gain measurements. One can see that gains of more that $10^5$ were easily achieved. One should note that at gains of more than $310^4$, photon feedback appeared, however at counting rated of less than $10^4$Hz it did not cause any special problems (the photon feedback pulses followed for several μs the main pulse and in the case of the integration time of the amplifier of 10 μs were counted as one pulse), so that one could operate the detector in photon counting mode until gains of $10^6$ and even more.
Fig. 7 shows the results of the measurements the stability in time or, to be more precise -the evolution of the QE and the gain with time. One should note that the detectors mentioned above were manufactured at different times. For example, the "oldest" one were the detectors with CsI photocathode and quartz windows; this is why we were able to monitor their stability over ~1, 5 year. The detectors with $MgF_2$ windows were manufactured only a few months ago and correspondingly in the Fig.7 the results obtained during this time interval are presented. One can conclude from this data that a rather good stability was achieved for all typed of detectors used in this work

## 4. Applications and Measurements
## 4-1.1 UV Flame and Spark Detectors.
### a) Indoor Applications

In many applications there is need to detect flame or sparks in daylight conditions, for example in houses, plants, storage buildings, etc. Recent studies show that it could be very efficiently done with UV sensitive detectors [14]. This is because most of the flames in air, almost independently on their nature, are emitting in the UV region (185-300 nm ) whereas the lamps used to illuminate rooms and buildings practically do not emit in this region due to the strong absorption in glass and other materials used in their design. The UV emission from the sun is also is fully blocked by the glass windows. Thus one can detect the UV light from flames with extremely high signal to noise ratio. There are some commercial detectors which explore this possibility: PMs combined with narrow band filters, solid state sensors and gaseous detectors. The cheapest are the gaseous detectors [14,15]. The example could be a Hamamatsu sealed gaseous detector with the metallic photocathode R2868 [12]. This detector is able to detect small flames in daylight conditions without using any filters. In our earlier preliminary experiments it was shown that much higher sensitivity for the flame detection in fully illuminated rooms could be achieved not with pure metallic photocathodes but with metallic ones covered by photosensitive layers, particularly with CsTe (see for example [16]). Unfortunately, CsTe photocathodes are difficult to manufacture and they remains stable only in the case of very clean gaseous conditions. As a consequence, the price of such sealed gaseous detector is as high as avacuum PM.

The fact that CsI photocathode could be exposed to air for 5-10 min without a strong loss in their QE dramatically simplifies the manufacturing of the sealed detector. Indeed, in contrast to vacuum photomultipliers where the photocathodes are manufacturing at the same time as the whole detector, the gaseous detectors with the CsI photocathodes could be manufactured in three steps as it was described above: 1) the coating of a metal cathode by an CsI layer, 2) assembling the whole detector in the air, 3) pumping, gas filling and sealing the detector. For example, one can suggest manufacturing the same detector as Hamamatsu R 2868, but with the cathode coated by the CsI layer. Since the assembling of the photocathode could be done in air, one can expect that the price of this modified device will no be increased by much.

One of the aims of this work was to compare the characteristics and the price of our sealed gaseous detectors to Hamamatsu R2868. In these comparative studies we used two designs of detectors with the CsI photocathodes: one with a quart and the other one with an $MgF_2$ window both filled with $Ar+10\%CO_2$ a total pressure of 1atm (see Fig. 2). Since Hamamatsu published some data for sensitivity of their devices to small flame flames produce by matches and cigarette lighters [12], we performed measurement with these flames (see Fig. 8). Some our results are presented in the Table 1. As one can se from the data presented, due to the CsI coating our detectors had remarkably high sensitivity to the UV radiation emitted by flames and at the same time practically were practically not sensitive to the visible light - the feature allowing them to operate without any problems in fully illuminated rooms. For example, a flame from a match or a cigarette lighter could be easily detected in a fully illuminated room at a distance of more than 30 m: the counting rate produced by the match was around 100 Hz, whereas the background counting rate (no match) in a fully illuminated room was below few Hz. The main conclusion from our measurements was that our detectors were 100- 1000 times more sensitive that the Hamamatsu one.

The other important advantage of our detectors is that in contrast to the Hamamatsu R2868 sensor our single- wire counters could operate in proportional modes and this allowed one to distictinguish between the signals produced by single photons and a few photons arriving almost at the same time. This in turn, allows reliable detection not only for quasy stationary flames, but also for the sparks- the feature which could be very important for creating a spark monitoring and spark alarm system.

### b) Outdoor Application

There is also a unique possibility to detect the UV light from the flames outside the building in the presence of strong sun light. This is based on the fact that the UV light from the sun with the wavelength of $\lambda < 280$ nm is fully adsorbed by the ozone in the upper layer of the atmosphere; however on ground level the atmosphere is transparent for this radiation. To be more precise: the atmosphere on the ground level is transparent for the radiation with wavelength of λ >185 nm at short distances from the UV source (up to 30-50 m) and transparent for the light with wavelength of λ > 240 nm for long distances (kilometers) from the source. This offers a unique possibility to detector the UV emission of flames (for example, forest fires) in the wavelength interval of 185-280 nm in direct sunlight conditions without any background from the sun.

The tests performed by us show that sealed detectors with CsI photocathodes are able to detect flames in outdoor conditions, however the background counting rate $N_{CsI}$ produced by the sunlight was rather high. This was because the QE of the CsI for λ> 280 nm is not at real zero (see Fig. 5) whereas the sun's emission in this spectral interval is extremely strong. As a result the convolution of the sunlight spectrum $S(\lambda)$ with the CsI QE ($Q_{CsI}$) in the spectral interval of 280-500 nm is not zero either:

$$N_{CsI} = \int_{280}^{500} S(\lambda) Q_{CsI}(\lambda) d\lambda > 0.$$

Measurements show that even in the presence of weak sunlight (a cloudy day) $N_{CsI} > 10^2$. Much better results were obtained with the photosensitive sealed detectors filled with EF vapors. As one can see from Fig. 5 the QE of the EF ($Q_{EF}$) for λ >200 nm is practically zero so one can expect that:

$$N_{EF} = \int_{280}^{500} S(\lambda) Q_{EF}(\lambda) d\lambda \sim 0.$$

Measurements performed with various flames (see Fig. 9) and at various conditions fully confirmed this expectation. As an example Fig.10 shows the counting rate vs. time (accumulated each 10 sec) for the detector filled by the EF and exposed to the direct sunlight. The mean number of the counts was 2.5 and it remained the same even if the detector was fully shielded from the light. So we attributed this small counting rate to the contribution from the cosmic radiation and from spurious pulses.

With such low background it was easy to detect small flames in the presence of the direct sun light. As an example in Fig. 10 shows the increase of the counting rate in the presence of a small flame placed 30 m away from the detector. Note that this was the flame produce by alcohol and it was practically invisible by human eye in sunlight conditions.

Comparison to commercial sensors of UV radiation from flames able to operate in sunlight conditions (see for example[14,15]) show that the sensitivity of our detector is at least 100 higher.

Note that our sealed gaseous detector are not only superior in sensitivity than commercially available devices but also considerably cheaper.

### 4-1.2. UV Detectors for Cryogenic TPCs.

Obviously, the sealed gaseous detector with CsI photocathodes could be used not only for the flame detections, but in many other applications- practically everywhere where detection of the UV light is needed. As an example we will present here the tests oriented on the application of our detectors for the noble liquid TPCs [8,17,18]. The key element in noble liquid TPCs are UV sensitive photodetectors which record the primary scintillation light from the noble liquids and generate the triggering signal to the readout system. In some designs, photodetectors are also supposed to detect the secondary scintillation light [17,19]. Currently expensive PMs are used for this purpose.
In our recent experiments [8,20-22] we have demonstrated that detectors with CsI photocathodes flushed by gas detectors are able to operate at cryogenic temperatures and detect with a high efficiency the scintillation lights from noble liquids. In spite of encouraging preliminary results, flushed by gas detectors have several inconvenient features, for example their working voltage V depends on the temperature. Certainly for the noble liquid TPC applications one would need stable detectors and preferably sealed detectors to avoid additional complications associated to the gas systems.

In this work for the first time we tested sealed gaseous detectors: single wire counters with $MgF_2$ windows filled with $Ar+10\%CH_4$ at p=1atm. Tests of these detectors were performed with the help of the cryostat described in [22]. As in work [20] inside the cryostat a "scintillation" chamber was installed - a gas chamber containing an $^{241}Am$ source and flushed, depending on the measurements, by Ar or by Xe gases at pressures of 1atm (Fig. 11). The alpha particles form the $^{241}Am$ produced the scintillation light which was detected simultaneously by our sealed gaseous detector and by a PM. If necessary, the scintillation chamber could be independently cooled to obtain a thin layer of LXe coating the surface of the $^{241}Am$ source.

Results obtained so far are very encouraging. As an example Fig.12 shows the gain and the QE measured as a function of temperature for the sealed gaseous detector. One can see that the gain remained constant and the QE degradeted only slightly. After the warming up the detector to room temperature, the QE values returned to their original ones.

Fig. 13 shows oscillogramms of signals from the sealed gaseous detector and from the PM simultaneously detecting scintillation the light from the LXe. One can see that the gaseous detector had much better signal to noise ratio.
The stability of our detectors was monitored at various temperature cycles for the period of more than three months; some results are presented in Fig. 7.

## 5. Conclusions

The main conclusion from our studies are that the developed by us technology allows to manufacture cost effective sealed gaseous detector having high quantum efficiency and good stability in time. Developed detectors are in fact gaseous photomultipliers which could be used in many other applications such as spectroscopy, biology, various security devices and so on. In contrast to commercially available vacuum PMs they are much simpler and cheaper and can operate in magnetic fields necessary in some measurements.
Note as well, that 1atm operation simplifies the feedthrough problems so that the detector could be easily made position sensitive( see for example[23,24]).

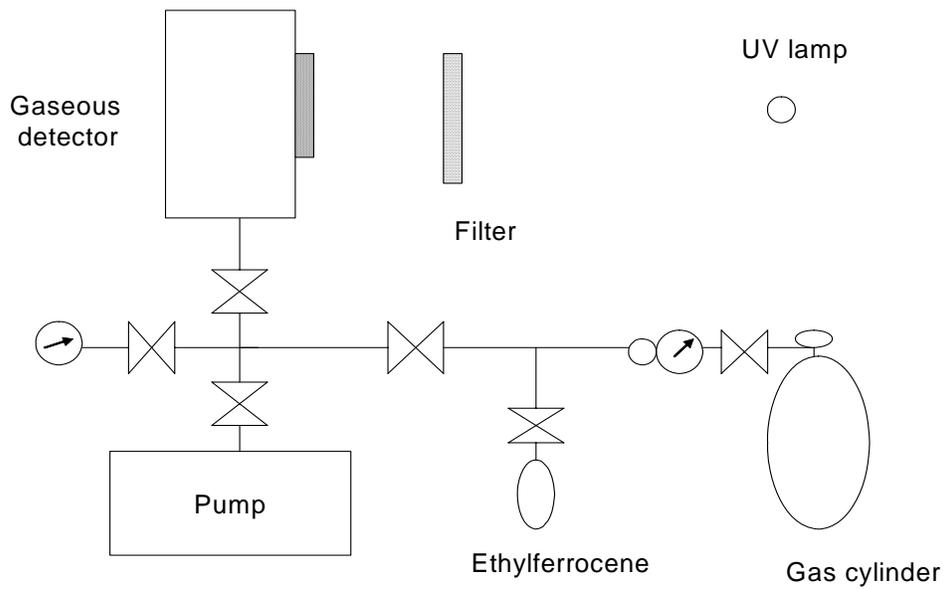

Fig.1. A schematic drawing of the set-up for manufacturing sealed gaseous detectors

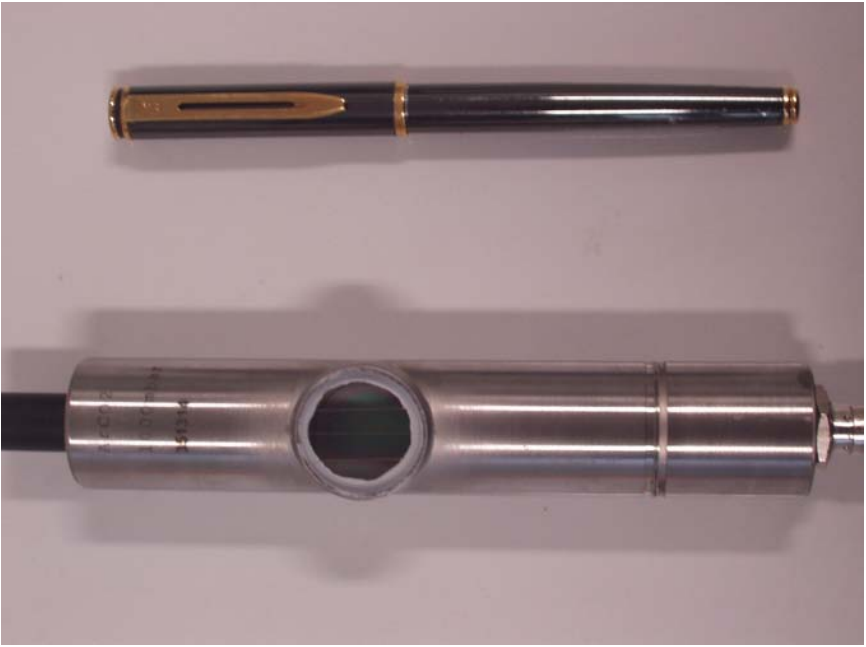

Fig. 2. Photo of the sealed gaseous detector with the CsI photocathode developed in the frame of collaboration between CERN and by Oxford Instruments

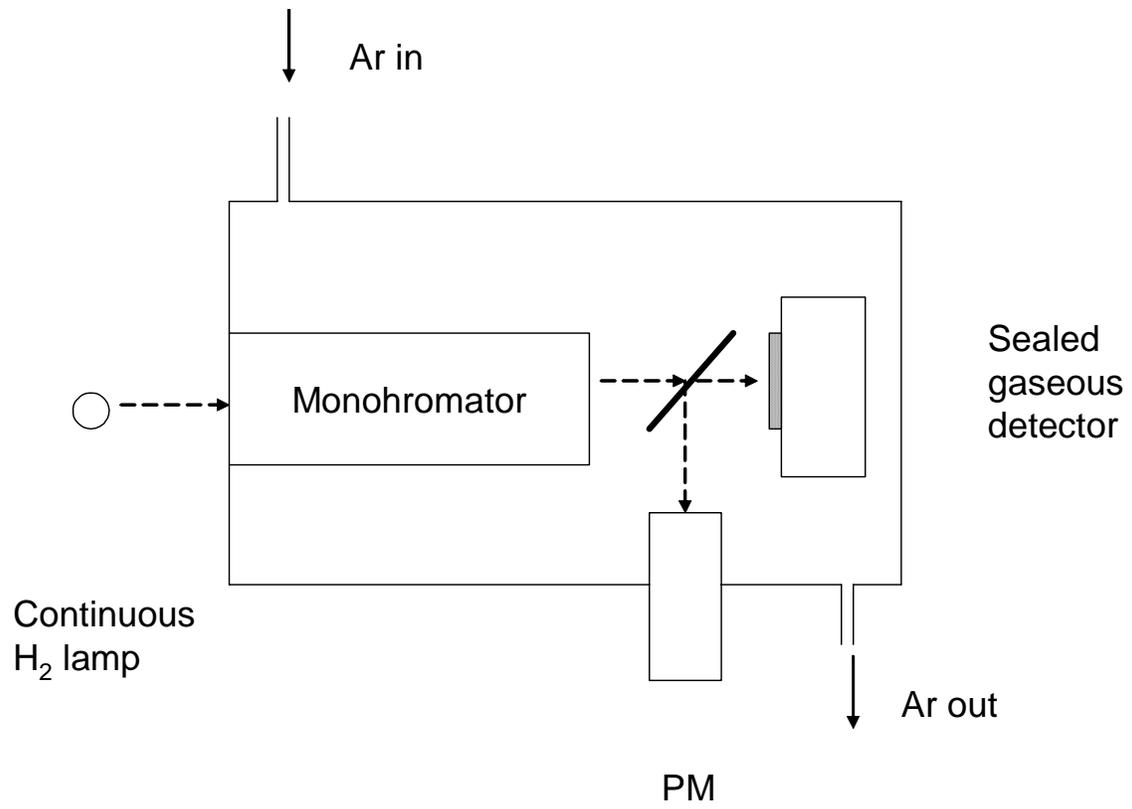

Fig.3. The set-up for measuring the QE of the detectors

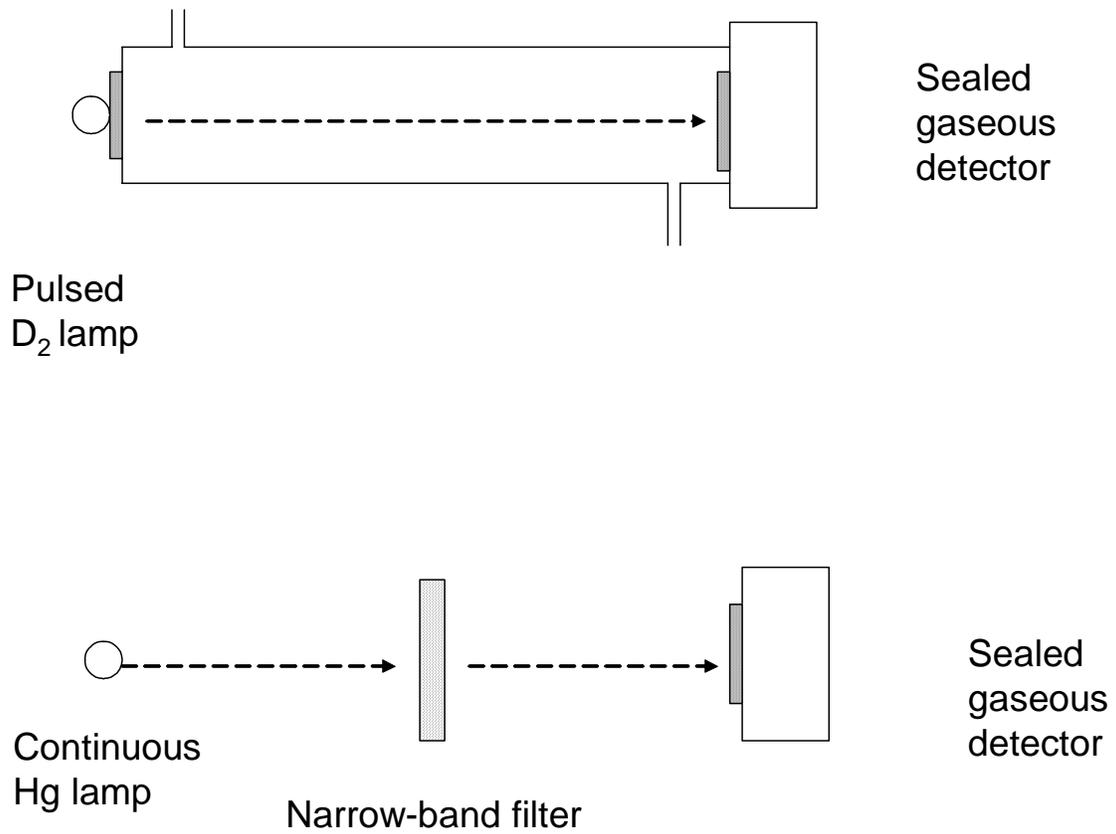

Fig.4. A schematic drawing of the set-up for monitoring the QE of the detectors and their gain stability in time

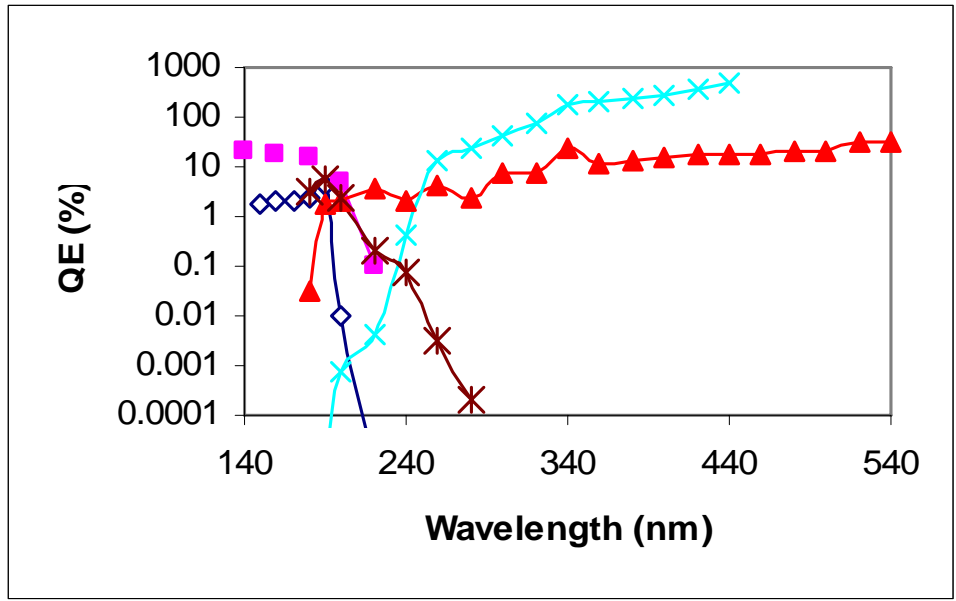

Fig.5. Results of the QE measurements for some our detectors: brown stars-the QE of the single –wire counter with the CsI photocathode and with the quartz window, rose squares –the QE of the gaseous detector with CsI photocathode and with the $MgF_2$ window, dark blue rhombus-the QE of the detector filled with the EF vapors ($MgF_2$ window).

In the same figure are presented the typical spectra of flames in air (red triangles) and the spectra of the sun (light blue crosses), both in arbitrary units. The latest data were taken from the Hamamtsu datasheet for the flame sensor R2868 [12]. Note that the curve for the spectra of the flames in air was verified in our measurement performed with help of the monohromator.

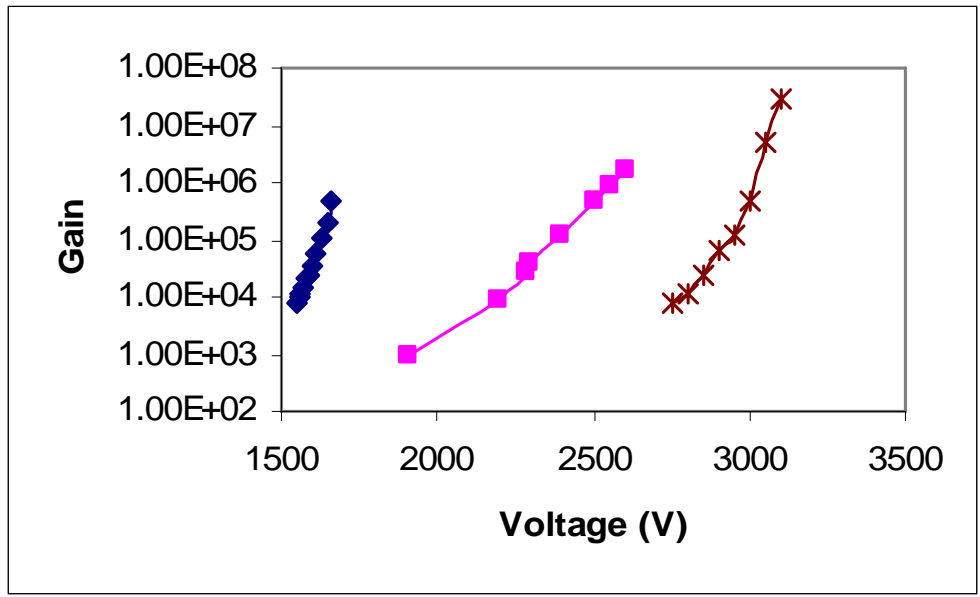

Fig. 6. Gain vs. voltage for some of our detectors: blue rhombus sealed single-wire counter with the CsI photocathode and with a diameter of the cathode cylinder of 20 mm (gas mixture Ar+10%$CO_2$), rose square- wire counter with the CsI photocathode with a diameter of the cathode cylinder of 30 mm(gas mixture Ar+10%$CH_4$), brown star-gain for the single wire- counter filled with EF vapors (gas mixture Ar+15%$CO_2$).

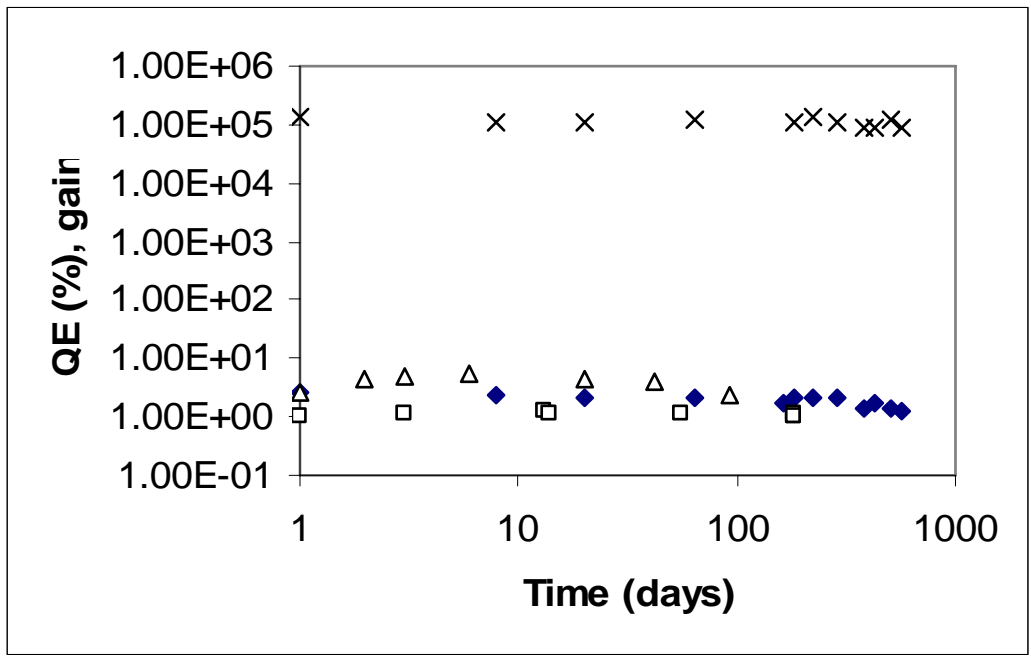

Fig.7. Results of monitoring the gas gain at the applied voltage of V=1670 V (upper curve-crosses ) and the QE for the period of more than 1,5 year for the detector with the CsI photocathode and the quartz window shown in (blue rhombus).
 Open triangles show the evolution of the QE with time for the gaseous detector with the CsI photocathode and with the MgF$_2$ window.
 Open squares show the results of the QE. monitoring for the detector filled with EF vapors.

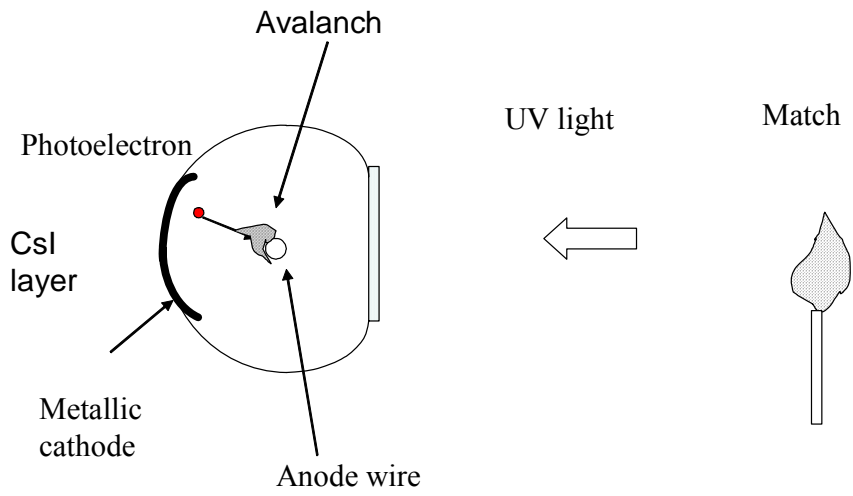

Fig. 8. A schematic drawing illustrating measurements performed with match

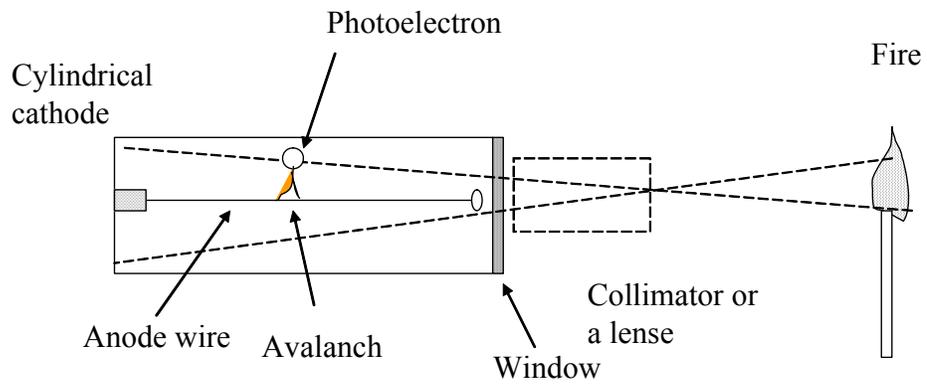

Fig. 9. A schematic drawing illustrating measurements performed with a sealed gaseous detector filled with EF vapors.

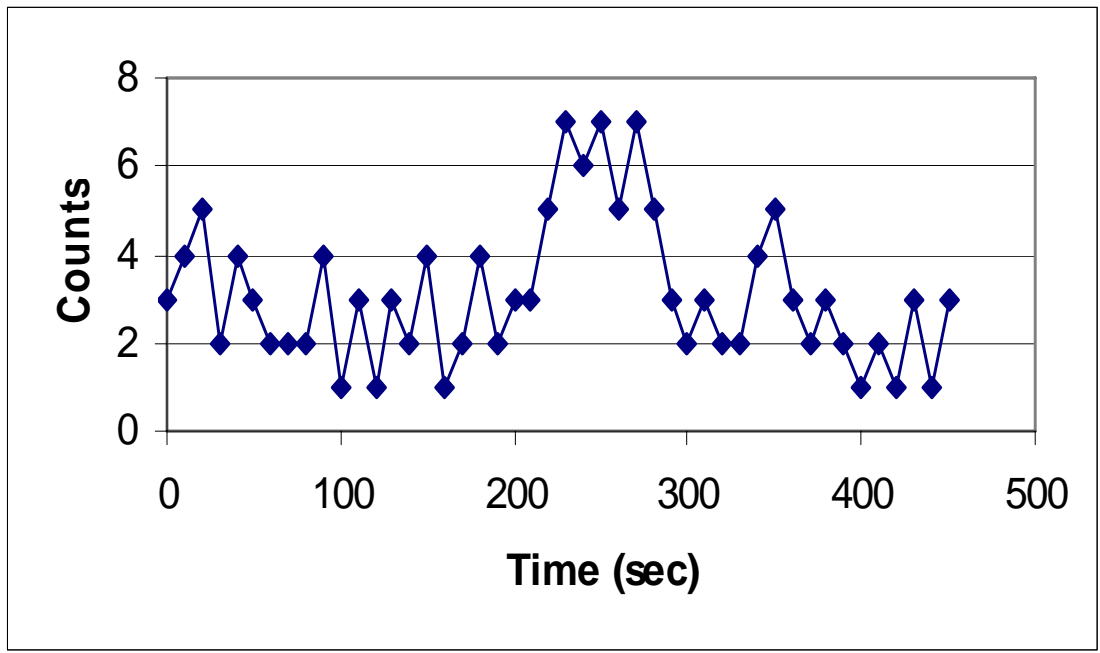

Fig. 10. Counts measured from the sealed gaseous detector filled with EF vapors (V=3 kV) in direct sunlight conditions. At the time interval 220-280 sec an alcohol fire (~5 x 5 x 5cm$^3$) was placed 30 m away from the detector. One can see that the counting rate almost doubled.

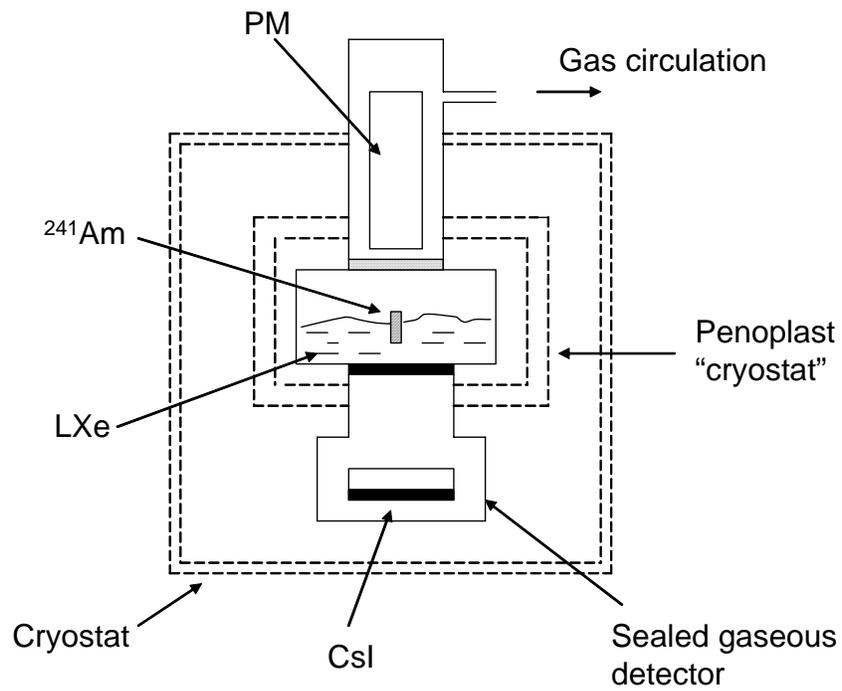

Fig. 11. A schematic drawing of the set-up for studies of operation of sealed gaseous detectors with CsI photocathodes at cryogenic temperatures

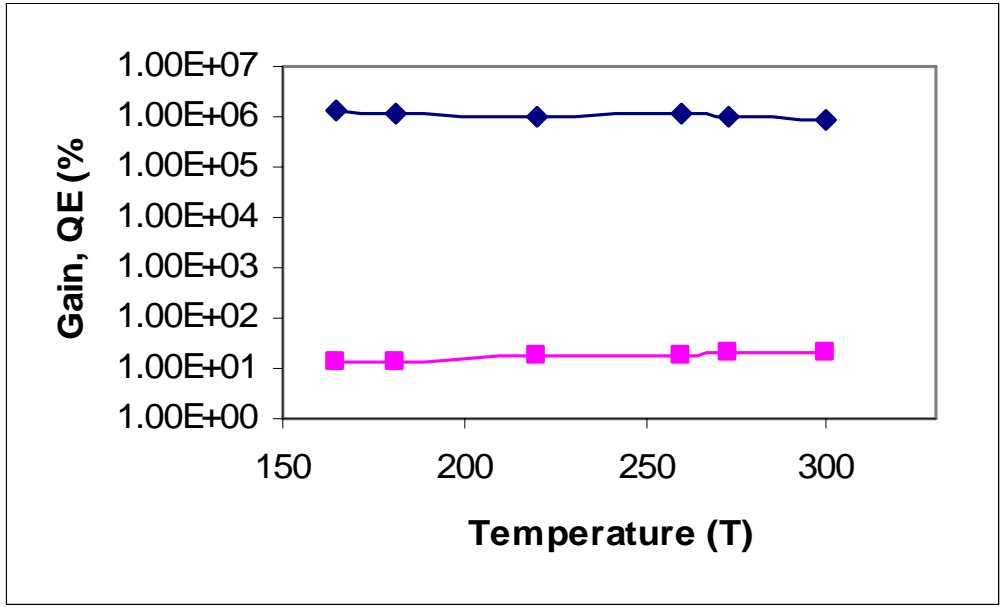

Fig.12. Gain at V=2500 V (the upper curve) and QE (lower curve) variations with the temperature for sealed gaseous detector with the CsI photocathode

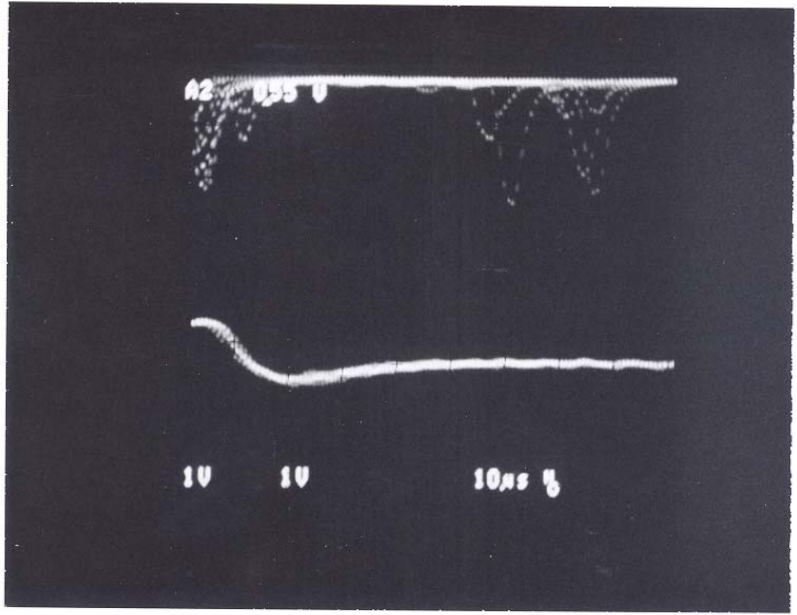

Fig.13. Oscillogramms of signals from the PM Schlumberger 541F-09-17 (upper curve) and from the sealed gaseous detector with CsI photocathode (lower curve) simultaneously detecting the scintillation light from LXe

| Hamamatsu R2868 | | Our detector with quartz window | Our detector ith MgF2 window |
| --- | --- | --- | --- |
| Distance (m) | Mean number of counts per 10sec | Mean number of counts per 10sec | Mean number of counts per 10sec |
| 1 | | 81579 | |
| 1,1 | 583 | | |
| 2,5 | 99 | | |
| 3 | 76 | 9015 | 87574 |
| 4,5 | 28 | | |
| 10 | 6 | 811 | 7902 |
| 20 | | | |
| 30 | | 92 | 876 |
| 60 | 0.1 | 19 | |

Tabl.1. Number of counts measured with three different sensors detecting the UV radiation from the match placed at different distances from the sensor's windows. One can conclude from these data that our sealed gaseous detector with the $MgF_2$ window is ~1152 more sensitive than the Hamamatsu R2868. The gaseous detector with the quartz window is~ 118 times more sensitive than the Hamamatsu one.